# Unlocking the Black Box: Analysing the EU Artificial Intelligence Act's Framework for Explainability in AI

Georgios PAVLIDIS[1]

**Abstract:** The lack of explainability of Artificial Intelligence (AI) is one of the first obstacles that the industry and regulators must overcome to mitigate the risks associated with the technology. The need for 'eXplainable AI' (XAI) is evident in fields where accountability, ethics and fairness are critical, such as healthcare, credit scoring, policing and the criminal justice system. At the EU level, the notion of explainability is one of the fundamental principles that underpin the AI Act, though the exact XAI techniques and requirements are still to be determined and tested in practice. This paper explores various approaches and techniques that promise to advance XAI, as well as the challenges of implementing the principle of explainability in AI governance and policies. Finally, the paper examines the integration of XAI into EU law, emphasising the issues of standard setting, oversight, and enforcement.

---

[1] Jean Monnet Chair and UNESCO Chair, Associate Professor of International and EU Law, NUP Cyprus, Director of the Jean Monnet Centre of Excellence AI-2-TRACE-CRIME (EU-funded), email: g.pavlidis@nup.ac.cy

# 1. The Imperative of Explainability: Navigating the Landscape of AI's Impact and Risks

Artificial intelligence (AI) has emerged as a fascinating and influential force in today's technological and business worlds. AI has already started to streamline mundane tasks, advance critical domains of scientific research and disrupt professions and industries. Whether in finance, commerce, healthcare or other fields, AI-based tools are steadily increasing their presence, thus permeating various organisational processes and decision-making.

As businesses realise the value of adopting AI, investment in the technology has surged, with global AI private investment reaching $91.9 billion in 2022, a sum 18 times greater than in 2013.[2] As a result, the global AI market, valued at $428 billion in 2022, is projected to grow to approximately $515.31 billion by 2023 and $2,025.12 billion by 2030.[3] AI tools have become essential for enterprises worldwide, with the percentage of companies using them doubling since 2017 and ranging between 50 and 60 percent in recent years. Moreover, the average number of AI capabilities (natural-language generation, robotic process automation, computer vision, etc.) employed in business units has also doubled, rising from 1.9 in 2018 to 3.8 in 2022.[4]

However, despite this excitement in the business world, serious concerns have arisen about the risks associated with AI. Historically, new technologies have often produced fear,[5] but many of the concerns linked to AI seem justified. Ethical dilemmas have become more pronounced as AI systems have started making decisions that affect human lives. This raises questions about accountability and transparency as well as the integration of these principles into law. At the societal level, there is the risk of bias and discrimination, as AI tools might inadvertently perpetuate existing inequalities.[6] Furthermore, the spectre of job displacement looms large, as automation and AI-driven processes threaten to reshape industries, increase labour market disparities and radically alter the employment landscape.[7] Security risks are coming to the forefront because the increasing reliance on AI creates new avenues for cyberattacks and data breaches.[8] Alongside these challenges, the concentration of power and data in the hands of a few dominant tech companies and entities raises concerns about the potential misuse of the technology in markets, public discourse and even politics.[9]

---

[2] Stanford University, 'Measuring trends in Artificial Intelligence', (2023) AI Index Report, <https://aiindex.stanford.edu/report/> accessed 15 January 2024.
[3] Fortune Business Insights, 'Artificial Intelligence Market' (2023) Market Research Report, <https://www.fortunebusinessinsights.com/industry-reports/artificial-intelligence-market-100114> accessed 15 January 2024.
[4] McKinsey, 'The state of AI in 2022—and a half decade in review' (2022) Survey, <https://www.mckinsey.com/capabilities/quantumblack/our-insights/the-state-of-ai-in-2022-and-a-half-decade-in-review> accessed 15 January 2024.
[5] Bernard Cohen, 'Commentary: The Fear and Distrust of Science in Historical Perspective' (1981) 6 Science, Technology, & Human Values 20; Marita Sturken, Douglas Thomas and Sandra Ball-Rokeach, *Technological Visions: Hopes and Fears That Shape New Technologies* (Temple University Press 2004).

We argue that the lack of explainability of AI is one of the first obstacles that the industry, regulators and supervisors must overcome to mitigate the above-mentioned risks. Usually, AI algorithms operate behind a veil of opacity, a situation also referred to as the 'black box problem'. In this situation, the inputs and outputs of an AI model may be known, but how exactly the inputs are transformed into outputs is difficult to determine.[10] AI opacity arises due to the complex architectures of AI models (particularly deep learning and neural networks), the high degree of non-linearity, the numerous layers and interconnected nodes, the high-dimensional data with their numerous features, the fact that learnt representations in AI models might not directly correspond to human-understandable features, and other factors. As a result, after the design and the learning environment of the AI model are established, the model determines the value of specific parameters and the way the answer is reached, which sometimes baffles its developers, especially when deep learning and neural networks are used.

---

[6] This may be due to several factors, such as biased training data, data collection methods, feature selection, and feedback loops; Frederik Zuiderveen Borgesius, 'Discrimination, Artificial Intelligence, and Algorithmic Decision-Making' (2018) Council of Europe Study.

[7] OECD, 'Artificial Intelligence and Employment' (2021) OECD Policy Brief; see also Accenture, 'A New Era of Generative AI for Everyone' (2023) Accenture Report. According to this report 40% of all working hours can be impacted by Large Language Models (LLMs) like GPT-4.

[8] European Union Agency for Cybersecurity, 'Artificial Intelligence Cybersecurity Challenges' (2020) ENISA Report.

[9] Nick Srnicek, 'Platform monopolies and the political economy of AI' in John McDonnell (ed) *Economics for the many* (Verso 2018); Pieter Verdegem, 'Dismantling AI capitalism: the commons as an alternative to the power concentration of Big Tech' (2022) AI & Society <https://doi.org/10.1007/s00146-022-01437-8> accessed 15 January 2024.

[10] Jenna Burrell, 'How the machine 'thinks': Understanding opacity in machine learning algorithms' (2016) 3 Big Data & Society 1; Davide Castelvecchi, 'Can we open the black box of AI?' (2016) 538 Nature 21; Warren von Eschenbach, 'Transparency and the Black Box Problem: Why We Do Not Trust AI' (2021) 34 Philosophy & Technology 1607.

The field of 'eXplainable AI' attempts to (XAI) bridge the gap between the opacity of machine decision-making and the human demand for comprehensibility. While transparency aims to unveil the inner workings of these systems, explainability goes beyond mere disclosure, revealing the cognitive processes that underlie decision-making. Therefore, ensuring XAI is more important and much more difficult, than ensuring transparency. Nevertheless, XAI diverges significantly from human explanation and justification in several key aspects, rooted in the disparities between machine learning algorithms and human cognitive processes. Unlike human cognition, AI operates through complex algorithms and neural networks, often yielding outputs that defy straightforward explication. Human explanations often draw upon personal experiences, background knowledge, and intuition, allowing for a narrative that encompasses context and emotional nuances. In contrast, AI algorithms process vast amounts of data and identify patterns that may elude human perception. Furthermore, human justifications often involve ethical, moral, or subjective considerations, while AI lacks inherent moral reasoning and operates based on learned patterns, potentially leading to decisions devoid of ethical dimensions. Finally, human justifications are also subject to interpretability and may be shaped by cultural, historical, or individual factors, contrary to the objectivity sought in AI systems.

The need for XAI is evident in fields where accountability, ethics and fairness are critical, such as healthcare, credit scoring, policing and the criminal justice system.[11] The opacity of AI models is the common denominator in most criticisms

---

[11] Roger Brownsword and Alon Harel, 'Law, liberty and technology: Criminal justice in the context of smart machines' (2019) 15 International Journal of Law in Context 107; Abdul Malek, 'Criminal courts' artificial intelligence: the way it reinforces bias and discrimination' (2022) 2 AI and Ethics 233; Michael Bücker and others, 'Transparency, auditability, and explainability of machine learning models in credit scoring' (2022) 73 Journal of the Operational Research Society 70; Georgios Pavlidis, 'Deploying artificial

of this new technology. For example, AI systems are criticised for generating unintended consequences, bias and violations of human rights, but it is the lack of explainability that hinders the identification, prevention, and mitigation of these problems.[12] In the absence of clear explanations of AI's underlying decision-making processes, the relevant authorities have no way to ensure compliance and supervision through regulations, regardless of when and how these will be established. Therefore, the principle of explainability is neither a simple software problem nor just an intellectually challenging quest for knowledge; it is a prerequisite for accountability, fairness, public trust, and effective regulation and supervision.

Recognising both the transformative potential and the significant risks of digitalisation, in particular AI, the European Union (EU) has taken ambitious steps to regulate digital innovation. Following the adoption of two landmark legislative instruments – the Markets in Crypto-Assets Regulation (MiCA)[13] and the Digital Operational Resilience Act (DORA)[14] – the next breakthrough at the EU level has been the initiative to adopt the Artificial Intelligence Act,[15] which aims to harness AI's potential while mitigating its dangers. The notion of explainability is one of the fundamental principles that underpin this legislative initiative, though the exact XAI techniques and requirements are still to be determined and tested in practice.

In the following sections, we will examine the philosophy, key provisions and intricacies of the EU AI Act (section 2); we will then delve deep into the principle of explainability, exploring various approaches and techniques that promise to

---

intelligence for anti-money laundering and asset recovery: the dawn of a new era' (2023) 26 Journal of Money Laundering Control 155.

[12] Executive Office of the U.S. President, 'Big data: a report on algorithmic systems, opportunity, and civil rights' (2016) Executive Office of the President Report.

[13] Regulation (EU) 2023/1114 of the European Parliament and of the Council of 31 May 2023 on markets in crypto-assets [2023] OJ L150/40; Georgios Pavlidis, 'Europe in the digital age: regulating digital finance without suffocating innovation' (2021) 13 Law, Innovation and Technology 464.

[14] Regulation (EU) 2022/2554 of the European Parliament and of the Council of 14 December 2022 on digital operational resilience for the financial sector [2022] OJ L333/1.

[15] European Commission, 'Proposal for a Regulation laying down harmonized rules on artificial intelligence (Artificial Intelligence Act)' (Communication) COM(2021) 206 final. The final text of the AI Act has been approved by the co-legislators (8 December 2023), but it has not been published yet in the Official Journal (January 2024).

advance XAI (section 3). We then shift our focus to the challenges of implementing the principle of explainability in AI governance and policies (section 4). Finally, we examine the integration of XAI into EU law (section 5), emphasising the issues of standard setting, oversight and enforcement.

## 2. The EU AI Act: Searching for Rules and Clarity

EU policymakers have consistently stressed the importance of fostering trustworthy, responsible and 'human-centric' innovation in the field of AI.[16] From the 2016 EU Global Strategy for Foreign and Security Policy[17] to the 2021 European Commission's plan for AI,[18] the EU has recognised the need to adapt current legislation, support global rules on AI and assume a leading role in the development of responsible AI.[19] The non-binding 2019 EU Ethics Guidelines for Trustworthy AI proposed the key principles in this context.[20] First, AI should empower humans and allow for human oversight through approaches such as human-in-the-loop, although there is no one-size-fits-all solution.[21] Second, AI systems must be technically robust and safe, and they must have contingency plans. Third, they should respect privacy, data protection and data governance. Fourth, transparency is crucial to ensure that stakeholders understand AI models and decisions. Fifth, AI systems must avoid unfair bias, promote diversity and sustainability and be accessible to all. Sixth, accountability mechanisms, such as auditability and redress options, must be deployed in responsible AI implementation. Soft-law principles for AI governance similar to the 2019 EU guidelines have been adopted by the Organisation for Economic Co-operation and Development (OECD) in 2019,[22] as well as by numerous other public and private organisations,[23] in an attempt to promote innovative yet responsible AI.

---

[16] Gonçalo Carriço, 'The EU and Artificial Intelligence: A Human-Centred Perspective' (2018) 17 European View 29; see also Paul Lukowicz, 'The Challenge of Human Centric AI' (2019) 3 Digitale Welt 9.
[17] European External Action Service, 'Shared Vision, Common Action: A Stronger Europe - A Global Strategy for the European Union's Foreign And Security Policy' (2016) <https://www.eeas.europa.eu/sites/default/files/eugs_review_web_0.pdf> accessed 15 January 2024.
[18] European Commission, 'Fostering a European approach to Artificial Intelligence' (Communication) COM(2021) 205 final.
[19] See also European Commission, 'Building Trust in Human-Centric Artificial Intelligence' (Communication) COM(2019) 168; European Commission, 'Fostering a European approach to Artificial Intelligence' (Communication) COM(2021) 205; European Commission, 'White Paper on Artificial Intelligence' (Communication) COM(2020) 65 final.
[20] European Commission, 'Ethics Guidelines for Trustworthy AI, High-Level Expert Group on AI' (2019) <https://digital-strategy.ec.europa.eu/en/library/ethics-guidelines-trustworthy-ai> accessed 15 January 2024.
[21] Lena Enqvist, 'Human oversight' in the EU artificial intelligence act: what, when and by whom?' (2023) Law, Innovation and Technology (Latest Articles), <https://doi.org/10.1080/17579961.2023.2245683> accessed 15 January 2024.

Recently, the soft-law approach to AI regulation seems to have made way for a stronger legislative approach. This has happened not only in the EU but also in other jurisdictions, with something of a 'race to AI regulation' apparently taking place.[24] The stronger approach advocates for the adoption of clear and binding rules for AI, with the support of effective enforcement mechanisms. The choice between hard and soft law has been the subject of scholarly interest in various areas of law and governance.[25] We argue that the shift from soft to hard law in AI regulation is timely and appropriate given the magnitude of the risks associated with this new technology. This does not mean that soft-law rules should be discarded completely;

---

[22] OECD, 'Recommendation of the Council on Artificial Intelligence' (2019) OECD/LEGAL/0449, <https://legalinstruments.oecd.org/en/instruments/oecd-legal-0449> accessed 15 January 2024.

[23] Ryan Budish, 'AI's Risky Business: Embracing Ambiguity in Managing the Risks of AI' (2021) 16 Journal of Business & Technology Law 259.

[24] Nathalie Smuha, 'From a 'race to AI' to a 'race to AI regulation': regulatory competition for artificial intelligence' (2021) 13 Law, Innovation and Technology 57; see the legislative initiatives on AI in Brazil (Projeto de Lei n° 2338, de 2023), in China (2021 regulation on recommendation algorithms; 2022 rules for deep synthesis; 2023 draft rules on generative AI), and in Canada (Draft law C-27, Digital Charter Implementation Act 2022, Part 3: Artificial Intelligence and Data Act). The United Kingdom does not plan to introduce sweeping new laws to govern AI, in contrast to the EU's AI Act, but to strengthen the roles of existing regulatory bodies like the Information Commissioner's Office, the Financial Conduct Authority, and the Competition and Markets Authority. These bodies will be empowered to provide guidance and oversee the use of AI within their respective areas of responsibility; UK Secretary of State for Science, 'Innovation and Technology, A pro-innovation approach to AI regulation' (2023) Policy Paper presented to the Parliament, <https://www.gov.uk/government/publications/ai-regulation-a-pro-innovation-approach/white-paper> accessed 15 January 2024.

[25] Among numerous studies, see Christine Chinkin, 'The Challenge of Soft Law: Development and Change in International Law' (1989) 38 International & Comparative Law Quarterly 850; Kenneth Abbott and Duncan Snidal, 'Hard and soft law in international governance' (2000) 54 International Organization 421; Bryan Druzin, 'Why does Soft Law Have any Power Anyway?' (2017) 7 Asian Journal of International Law 361.

they offer the advantages of flexibility and adaptability, and they can be employed to support hard law as mutually reinforcing complements.[26] A set of enforceable rules, the violation of which would lead to legal consequences and penalties, would provide a stronger deterrent against non-compliance and help mitigate the dangers of AI more effectively.[27] At the EU level, a clear legal framework for AI, applied uniformly across all member states, would reduce ambiguity and ensure that the rights and responsibilities of AI developers and users are interpreted in a standardised manner in the EU single market. A binding legal framework for AI would also enhance accountability and transparency by requiring entities in all EU member states to disclose their actions and practices to the relevant authorities, thus allowing supervisors to evaluate compliance and potential violations. These elements promise to instil greater confidence in businesses, investors and consumers, thereby increasing engagement with AI technologies; they would also effectively address risks and prevent forum shopping in the single market.

To make this happen, the EU must grapple with the challenge of balancing innovation and responsible AI when designing and implementing a new binding framework. In April 2021, the European Commission took on this challenge by putting forward an ambitious proposal for an EU regulatory framework concerning AI based on Articles 114 and 161 of the Treaty on the Functioning of the European Union. Following the approval of the European Council's general position in December 2021 and the European Parliament's vote in June 2023, EU lawmakers have entered into negotiations (trialogue) and finalised the text in December 2023.[28] These discussions entailed several amendments to the European Commission's initial proposal, including revisions to the definition of AI systems and to the list of prohibited AI systems.

---

[26] Gregory Shaffer and Mark Pollack, 'Hard Versus Soft Law in International Security' (2011) 52 Boston College Law Review 1147.
[27] Emer O'Hagan, 'Too soft to handle? A reflection on soft law in Europe and accession states' (2004) 26 Journal of European Integration 379; Jan Klabbers, 'The Undesirability of Soft Law' (1998) 67 Nordic Journal of International Law 381.
[28] On these negotiations, see European Parliament Legislative Observatory, 'Artificial Intelligence Act, 2021/0106(COD)' (2023) <https://oeil.secure.europarl.europa.eu/oeil/popups/ficheprocedure.do?reference=2021/0106(COD)&l=en> accessed 15 January 2024.

We argue that the decision to introduce a new instrument was appropriate and opportune. While existing EU rules, especially the General Data Protection Regulation (GDPR), do provide some guidance for safeguarding data in the realm of AI, they fail to directly address several data protection challenges associated with this technology. A 2020 study by the European Parliament identified the shortcomings of the GDPR in terms of the processing of personal data enabled by AI, including the lack of concrete obligations on XAI.[29] A new harmonised framework would have the benefit of comprehensively addressing the development of AI in the EU single market, as well as the use of AI products and services. Furthermore, to ensure effectiveness and prevent forum shopping, the new regulatory framework should cover users of AI systems located in the EU and providers of such systems established both in the EU and in a third country if they place their products or services in the EU market.[30]

After choosing to put regulations in place, the next important question is: What exactly will be subjected to regulatory measures? In its initial proposal for the EU AI Act, the European Commission put forward a technology-neutral, albeit quite broad, definition of AI systems.[31] This definition has faced criticism for its extensive scope, leading the EU Council to propose a more refined description with clearer criteria to distinguish AI from established software systems.[32] Conversely, the

---

[29] According to the study, 'it should be made clear that controllers have best-effort obligations to provide data subjects with individualised explanations when their data are used for automated decision-making: these explanations should specify what factors have determined unfavourable assessments or decisions […] This obligation has to be balanced with the need to use the most effective technologies. Explanations may be high-level, but they should still enable users to contest detrimental outcomes'; European Parliament, 'The impact of the GDPR on artificial intelligence' (2020) Scientific Foresight Unit (STOA) Options Brief, PE 641.530.

[30] This is an illustration of the concept known as the 'Brussels effect', which pertains to the EU's independent ability to control global markets through regulations; see Anu Bradford, *The Brussels Effect – How the European Union rules the world* (Oxford University Press 2020).

[31] Article 3(1) of the proposal defined AI system as 'software that is developed with [specific] techniques and approaches [listed in Annex 1] and can, for a given set of human-defined objectives, generate outputs such as content, predictions, recommendations, or decisions influencing the environments they interact with'.

[32] Council of the European Union, 'Proposal for a Regulation of the European Parliament and of the Council laying down harmonised rules on artificial intelligence (Artificial Intelligence Act) and amending certain Union legislative acts - General approach' Doc. 14954/22, 25 November 2022; according to the Council's definition, AI systems are

European Parliament proposed to amend the definition to align it with the OECD one.[33] Ultimately, this was the approach adopted by European Parliament and Council of the EU in the political agreement reached in December 2023.[34] Nevertheless, no agreed definition of AI has been given at the level of the EU-U.S. Terminology and Taxonomy for AI initiative.[35] In this context, it has been correctly pointed out that rather than relying on the definition of the term 'AI,' policymakers should focus on identifying the specific risks they want to reduce.[36]

This brings us to the next challenge for the EU AI Act – that is, the classification of the risks that are associated with specific uses of the technology. The classification system in the EU AI Act relies on a 'risk-based approach' that assigns different requirements and obligations to each risk category.[37] Risk-based regulation, which has become somewhat of a buzzword in the past few decades,[38] has proved to be effective in areas such as banking and the fight against money laundering.[39] Applying this approach in the AI context holds substantial promise; it

---

'systems developed through machine learning approaches and logic- and knowledge-based approaches'.

[33] European Parliament, 'Amendments adopted on 14 June 2023 on the proposal for a regulation of the European Parliament and of the Council on laying down harmonised rules on artificial intelligence (Artificial Intelligence Act) and amending certain Union legislative acts' COM(2021)0206 – C9-0146/2021 – 2021/0106(COD); according to the European Parliament's proposal, 'artificial intelligence system' (AI system) means a machine-based system that is designed to operate with varying levels of autonomy and that can, for explicit or implicit objectives, generate outputs such as predictions, recommendations, or decisions, that influence physical or virtual environments'.

[34] The final text of the AI Act has not been published yet in the Official Journal (15 January 2023).

[35] Trade and Technology Council, 'EU-U.S. Terminology and Taxonomy for Artificial Intelligence' (2023) Annex A <https://digital-strategy.ec.europa.eu/en/library/eu-us-terminology-and-taxonomy-artificial-intelligence> accessed 15 January 2024.

[36] Jonas Schuett, 'Defining the scope of AI regulations' (2023) 15 Law, Innovation and Technology 60.

[37] Michael Veale and Frederik Zuiderveen Borgesius, 'Demystifying the Draft EU Artificial Intelligence Act' (2021) 4 Computer Law Review International 97.

[38] Risk-based regulation is no panacea; serious issues arise regarding justification and legitimation, along with risk-scoring, enforcement, and compliance; Robert Baldwin, Martin Cave and Martin Lodge, 'Risk-based Regulation' in Robert Baldwin and others (eds), *Understanding Regulation: Theory, Strategy, and Practice* (Oxford University Press 2011).

[39] The key international standard-setter in this field, the Financial Action Task Force, summarized the advantages of this approach: 'A risk-based approach involves

also builds on the principle of proportionality, since regulations apply to AI applications only to the extent required to handle the specific level of risk they pose. More specifically, AI systems deemed to pose 'unacceptable' risks would be banned. A range of 'high-risk' AI systems would be permitted only if they met specific requirements and obligations before entering the EU single market. Limited transparency obligations would apply to systems that present limited risks, such as the need to indicate that an AI system is being used and interacts with humans. In this model, the provision of information and transparency would be mandatory for high-risk AI systems. As we will see in the following section, the challenge here is to decide which criteria and methods of explainability should be required by an AI regulation.

### 3. Unravelling the Magic: Approaches to and Techniques for XAI

The principle of explainability and the field of XAI are often put forward as the solution to the above-mentioned 'black box problem'. Although XAI lacks a commonly agreed definition, it can be broadly described as AI techniques that enable human users to comprehend, trust and manage AI effectively. In the literature, 'explainability' is often used interchangeably with 'interpretability'.[40] Explainability differs from simple transparency, a notion that refers to the degree to

---

    tailoring the supervisory response to fit the assessed risks. This approach allows supervisors to allocate finite resources to effectively mitigate the [...] risks they have identified and that are aligned with national priorities [...] A robust risk-based approach includes appropriate strategies to address the full spectrum of risks, from higher to lower risk sectors and entities. Implemented properly, a risk-based approach is more responsive, less burdensome, and delegates more decisions to the people best placed to make them'; Financial Action Task Force, *Risk-Based Supervision* (FATF 2021).

[40] Christian Meske and others, 'Explainable Artificial Intelligence: Objectives, Stakeholders, and Future Research Opportunities' (2022) 39 Information Systems Management 53; these authors propose the following method to distinguish between the two concepts: when humans can comprehend the system's logic and behaviors directly, without supplementary clarifications, the right term to use is 'interpretable AI'. This might be perceived as an inherent trait of the system. However, if humans necessitate explanations as an intermediary to fathom the system's procedures, the field is termed as research on 'explainable AI'.

which an AI system is open and observable to humans.[41] Transparency means that humans have access to critical information, such as the data used to train the AI system, the manner in which the data are gathered and stored, and who has access to the data the system collects. Still, making information available is not equivalent to making it comprehensible, which is the purpose of explainability. The importance of XAI has been recognised not only by the EU but also by key global players, such as the US Department of Defence,[42] big tech companies,[43] international organisations,[44] smaller companies and start-ups.[45] The question is how explainability can be defined and implemented in an optimal way in order to foster trust, fairness and accountability.

XAI techniques can provide different types of explanations – local or global – which explain individual predictions or offer insights into the model's behaviour, respectively. Several classification models and taxonomies for XAI methods have been proposed.[46] Model architectures that are designed with interpretability in mind may use decision trees, linear models and other mechanisms, and their performance, limitations and applicability may vary.[47] Without getting into too much technical detail, it is worth highlighting the variety of methods used in XAI, which is the key challenge when it comes to integrating this notion into law.

---

[41] OECD, 'Recommendation of the Council on Artificial Intelligence' (2019) OECD/LEGAL/0449, <https://legalinstruments.oecd.org/en/instruments/oecd-legal-0449> accessed 15 January 2024 (Principle 1.3 – Rationale).

[42] On the XAI initiatives of the U.S. Department of Defense (Defense Advanced Research Projects Agency, DARPA) see David Gunning and others, 'DARPA's explainable AI (XAI) program: A retrospective' (2021) 2 Applied AI Letters, <https://doi.org/10.1002/ail2.61> accessed 15 January 2024.

[43] For example, IBM has proposed a precision regulation framework, which also refers to the need for explainable AI: 'Any AI system on the market that is making determinations or recommendations with potentially significant implications for individuals should be able to explain and contextualize how and why it arrived at a particular conclusion. To achieve that, it is necessary for organizations to maintain audit trails surrounding their input and training data. Owners and operators of these systems should also make available — as appropriate and in a context that the relevant end-user can understand — documentation that detail essential information for consumers to be aware of, such as confidence measures, levels of procedural regularity, and error analysis'; see <https://www.ibm.com/policy/ai-precision-regulation/> accessed 15 January 2024.

[44] See e.g. the Global Summit 'AI for Good', organized by the International Telecommunications Union, in partnership with 40 UN Agencies in 2023; <https://aiforgood.itu.int/about-ai-for-good/> accessed 15 January 2024.

[45] In this context, the problem of inequity of access to XAI emerges; due to the costs of XAI, SMEs may be get left behind in this process and be priced out of XAI; see Jonathan Dodge, 'Position: Who Gets to Harness (X)AI? For Billion-Dollar Organizations Only' (2021) Joint Proceedings of the ACM IUI 2021 Workshops, <https://ceur-ws.org/Vol-2903/IUI21WS-TExSS-5.pdf> accessed 15 January 2024.

[46] Riccardo Guidotti and others, 'A Survey of Methods for Explaining Black Box Models' (2018) 51 ACM Computing Surveys 1.

XAI can take a quantitative route, which aims to provide insights into the importance of certain features and establish suitable scaling and normalisation methods for them.[48] Because minimal feature changes in AI models can lead to different model predictions, several techniques attempt to determine the importance of features.[49] A closely related approach to XAI focuses on feature interactions, i.e. how combinations of features influence the model's output.[50] The objective here is to uncover intricate relationships in the data; however, this requires sophisticated visualisation techniques, as we will see in the next section. Several counterfactual explanation methods go beyond correlation and explore causality by examining what changes in the inputs would lead to different outcomes.[51] XAI may also employ sensitivity analysis, in which input features are varied systematically to observe the corresponding changes in predictions, thereby identifying critical features. Doing so is computationally intensive when there are multiple features and high-dimensional data. Furthermore, XAI may go beyond deterministic explanations and employ probabilistic models to estimate uncertainty and prevent potentially risky predictions, especially when scenarios demand risk assessments.[52] Finally, the temporal dimension is crucial, and many XAI methods are designed to understand how decisions unfold over time, how patterns change, and how time intervals affect accuracy, especially in fast-changing situations.[53]

---

[47] Alex Freitas, 'Comprehensible classification models: A position paper' (2013) 15 ACM SIGKDD Explorations Newsletter 1.
[48] An-phi Nguyen and María Rodríguez Martínez, 'On Quantitative Aspects of Model Interpretability' (2020) ArXiv, <https://arxiv.org/abs/2007.07584> accessed 15 January 2024; B Kim and others, 'Interpretability Beyond Feature Attribution: Quantitative Testing with Concept Activation Vectors (TCAV)' (2018) 80 Proceedings of the 35th International Conference on Machine Learning 2668.
[49] E.g. permutation importance, SHAP (SHapley Additive exPlanations), and LIME (Local Interpretable Model-agnostic Explanations).
[50] Sejong Oh, 'Predictive case-based feature importance and interaction' (2022) 593 Information Sciences 155.
[51] Riccardo Guidotti, 'Counterfactual explanations and how to find them: literature review and benchmarking' (2022) Data Mining and Knowledge Discovery, <https://doi.org/10.1007/s10618-022-00831-6> accessed 15 January 2024.
[52] Riccardo Guidotti and others, 'A Survey of Methods for Explaining Black Box Models' (2018) 51 ACM Computing Surveys 1.

## 4. Challenges and Considerations regarding XAI Implementation

While the importance of explainability is widely acknowledged and despite years of research in the field, implementing explainability has proved challenging due to issues of complexity and scalability.

As AI models grow in size and complexity, so do the practical obstacles to producing XAI models that remain understandable and trustworthy. Another factor that should be considered is the cost of XAI, which rises with the increasing complexity of the technology. A major obstacle is the lack of agreement on how to perform an objective evaluation of XAI methods. Moreover, if explainability requirements vary by jurisdiction, businesses must develop different algorithms and explainability methodologies for different markets, which would increase their costs and put small and medium-sized enterprises in an unfavourable position. There is also an inherent trade-off between model transparency and performance, which may lead different industries and AI applications to prioritise one over the other. Certain categories of AI systems are more prone to experiencing a decline in accuracy and performance if they are mandated to be explainable.[54] This is because the process of achieving explainability can involve simplifying the solution variables to the level where human comprehension is possible. Obviously, this approach might not be optimal for intricate, multidimensional problems.

Ethical and privacy considerations also arise in the context of XAI. First, providing explanations for ethical aspects of AI decision-making requires a prior and clear definition of fairness and bias. However, this definition might be subjective and highly dependent on context, which might lead to differing interpretations.

---

[53] Thomas Rojat and others, 'Explainable Artificial Intelligence (XAI) on Time Series Data: A Survey' (2021) ArXiv, <https://arxiv.org/abs/2104.00950> accessed 15 January 2024.

[54] OECD, 'Recommendation of the Council on Artificial Intelligence' (2019) OECD/LEGAL/0449, <https://legalinstruments.oecd.org/en/instruments/oecd-legal-0449> accessed 15 January 2024 (Principle 1.3 – Rationale).

Legal definitions of bias may also differ from one jurisdiction to another, which further complicates things. Second, XAI must take into account privacy concerns because explanations of model decisions must not disclose sensitive information, but they must still be able to provide meaningful insights. As is the case with bias, the legal requirements for privacy may differ across jurisdictions.

Furthermore, XAI must balance the need for explanations with that to defend proprietary algorithms, as source code and datasets may be subject to intellectual property or trade secrets. Therefore, increased transparency obligations imposed by law should not disproportionately affect the intellectual property rights of AI developers. In the context of the EU AI Act, transparency requirements encompass the essential details needed for individuals to exercise their right to an effective remedy and for supervision and enforcement authorities to exercise their mandates.[55] Also, any divulgence of information must adhere to relevant legal frameworks, including Directive 2016/943,[56] which safeguards trade secrets from unlawful acquisition, use and disclosure. Therefore, when supervision and enforcement authorities require access to confidential information or source code in order to assess compliance, they are bound by strict confidentiality obligations.

Finally, human-AI interaction plays a key role in XAI by highlighting methods for effectively communicating complex AI concepts to stakeholders, thus fostering transparency and trust.[57] This issue poses several challenges. First, algorithms may be complex and non-linear;[58] therefore, the explanations provided by them may not align with human intuition, which undermines trust in the AI system. Second, there is a trade-off between accurate representation and simplification; a balance must be found to avoid oversimplification and the loss of essential information. Indeed, the complexity of AI models makes it difficult to design visual representations that ensure user understanding and effectively communicate intricate model behaviours without overwhelming users or leaving out essential information.[59] Third, when AI systems make biased or unfair decisions, it is necessary to ensure that the explanations given are not just post hoc justifications. Fourth, clarifications should be user-centric – that is, they should take into account the varying expertise levels of different user groups.[60] Explanations must be tailored to the technical proficiency of users, who range from laypersons to experts (e.g., oversight authorities), and they must be meaningful and useful to each of them. This also means that clarifications must be relevant and informative for the decision-making processes of each group of users, without imposing an excessive cognitive load. This could be ensured by enabling interaction between humans and AI in the form of feedback loops. Users' feedback on clarifications could be employed to refine the explanation methods over time, thus factoring in the evolution of user preferences.[61]

---

[55] European Commission, 'Explanatory Memorandum, Proposal for a Regulation laying down harmonized rules on artificial intelligence (Artificial Intelligence Act)' (Communication) COM(2021) 206 final, section 3.5.

[56] Directive (EU) 2016/943 of the European Parliament and of the Council of 8 June 2016 on the protection of undisclosed know-how and business information (trade secrets) against their unlawful acquisition, use and disclosure [2016] OJ L157/1.

[57] Andrew Silva and others, 'Explainable Artificial Intelligence: Evaluating the Objective and Subjective Impacts of XAI on Human-Agent Interaction' (2023) 39 International Journal of Human–Computer Interaction 1390; Michael Chromik and Andreas Butz, 'Human-XAI Interaction: A Review and Design Principles for Explanation User Interfaces' in C Ardito and others, *Human-Computer Interaction – INTERACT 2021* (Springer 2021).

[58] Elena Benderskaya, 'Nonlinear Trends in Modern Artificial Intelligence: A New Perspective' in Jozef Kelemen, Jan Romportl and Eva Zackova (eds), *Beyond Artificial Intelligence. Topics in Intelligent Engineering and Informatics* (Springer 2013); Marina

## 5. Integrating AI Explainability into EU Law

Given these challenges, it is worth examining how to harmoniously meld the principle of AI explainability with the current legal landscape of the EU. As already mentioned, the EU must first determine which AI systems qualify as high risk and thus fall under the obligation of explainability. The EU must also specify what exactly the obligation of explainability would entail; in other words, it must spell out which

---

Vidovicet and others, 'Feature Importance Measure for Non-linear Learning Algorithms' (2016) ArXiv, abs/1611.07567 accessed 15 January 2024.

[59] Antoine Hudon and others, 'Explainable Artificial Intelligence (XAI): How the Visualization of AI Predictions Affects User Cognitive Load and Confidence' in Fred Davis and others (eds), *Information Systems and Neuroscience* (Springer 2021).

[60] Maximilian Förster and others, 'User-centric explainable AI: design and evaluation of an approach to generate coherent counterfactual explanations for structured data' (2022) Journal of Decision Systems (Latest Articles), <https://doi.org/10.1080/12460125.2022.2119707> accessed 15 January 2024.

[61] Rui Zhang and others, 'An Ideal Human: Expectations of AI Teammates in Human-AI Teaming' (2021) 4 Proceedings of the ACM on Human-Computer Interaction 1.

XAI methods and standards should be met to ensure compliance. Evidently, EU law should introduce a distinction between different levels of explainability and detail the technical aspects of implementing XAI. It must also address the intersection of XAI and privacy as well as consider the need to provide explanations to users in compliance with data protection regulations, such as the GDPR.[62]

To do so effectively, the EU has opted for the method of co-regulation through standardisation based on the New Legislative Framework (NLF).[63] The EU AI Act recognises the key role that standardisation should play in providing technical solutions to providers and in ensuring compliance with the new framework for AI, which will include the chosen standards for explainability.[64] Some have argued that entrusting the creation of regulations to organisations operating under private legal frameworks, such as CENELEC,[65] may be problematic at the level of judicial scrutiny.[66] Nevertheless, amid the complexity of AI advancements, novel governance models (e.g., hybrid governance) will inevitably gain traction. This type of governance merges state and non-state actors as well as the public and private realms through varying degrees of regulation, such as co-regulation and self-regulation.[67] These models, which are complementary to traditional command-and-control legislation,[68] can identify the changing risks of evolving AI systems more effectively, thus prompting the public and different stakeholders to create the necessary norms.

---

[62] Regulation (EU) 2016/679 of the European Parliament and of the Council of 27 April 2016 on the protection of natural persons with regard to the processing of personal data and on the free movement of such data, and repealing Directive 95/46/EC [2016] OJ L119/1.

[63] The new legislative framework, adopted in 2008, deals with conditions for the placement of products in the EU internal market. The framework clarifies rules for accreditation and CE marking and provides a toolbox for consistent sector-specific legislation; according to the 2022 evaluation, the NLF has achieved most of these objectives; see European Commission, 'Executive Summary of the Evaluation of the New Legislative Framework' (Commission Staff Working Document) SWD(2022) 365 final SWD(2022) 364 final

[64] Recital (61) EU AI Act; see also Martin Ebers, 'Standardizing AI - The Case of the European Commission's Proposal for an Artificial Intelligence Act' in Larry DiMatteo, Michel Cannarsa and Cristina Poncibò (eds), *The Cambridge Handbook of Artificial Intelligence: Global Perspectives on Law and Ethics* (Cambridge University Press 2021).

[65] The European Committee for Electrotechnical Standardization (CENELEC) is an association that brings together the National Electrotechnical Committees of 34 European countries, which work together to prepare voluntary standards in the electrotechnical field.

[66] Michael Veale and Frederik Zuiderveen Borgesius (2021), 'Demystifying the Draft EU Artificial Intelligence Act' (2021) 4 Computer Law Review International 97, citing the Case C-171/11 *Fra.bo SpA v Deutsche Vereinigung des Gas- und Wasserfaches eV* [2012] ECLI:EU:C:2012:453, as well as the opinion of Advocate General Trstenjak in the same case.

More specifically, under the EU AI Act, the compliance and enforcement system for stand-alone, high-risk AI models (Annex III) is modelled after the NLF and combines ex ante conformity assessments (internal checks) with ex post enforcement. Ex ante assessment of stand-alone, high-risk AI systems (i.e., systems that are new or represent significant modifications of existing systems) requires compliance with key obligations, such as explainability, as well as robust quality and risk management systems and post-market monitoring. Following this assessment, providers must register high-risk AI models in an EU database managed by the European Commission to enhance public transparency, oversight, and ex post supervision by the authorities. In this context, EU law must establish effective accountability mechanisms[69] to make developers, operators and users of AI systems comply with the EU AI Act and its XAI components. This will involve establishing clear lines of responsibility, requiring documentation of the AI development process, and setting up audit trails for AI-generated decisions. This is exactly what the EU AI Act's provisions concerning technical documentation, the EU declaration of conformity, and the conformity assessment procedures aim to achieve. Naturally, it will be essential to have effective, proportionate, and deterrent penalties, including administrative fines, for violations, as well as incentives for compliance.[70]

---

[67] Araz Taeihagh, 'Governance of Artificial Intelligence' (2021) 40 Policy and Society 137.

[68] Linda Senden, 'Soft Law, Self-Regulation and Co-Regulation in European Law: Where Do They Meet?' (2005) 9 Electronic Journal of Comparative Law, <https://ssrn.com/abstract=943063> accessed 15 January 2024. In this context, technological management must also be part of the regulatory framework; see Roger Brownsword, *Law, Technology and Society: Reimagining the Regulatory Environment* (Routledge Taylor & Francis Group 2019).

[69] Claudio Novelli, Mariarosaria Taddeo and Luciano Floridi, 'Accountability in Artifcial Intelligence: what it is and how it works' (2023) AI & Society, <https://doi.org/10.1007/s00146-023-01635-y > accessed 15 January 2024.

[70] In this context, the Commission's proposal moves in the right direction; for some types of serious infringements, it provides for administrative fines of up to 30 000 000 EUR or, if the offender is company, up to 6 % of its total worldwide annual turnover for the preceding financial year, whichever is higher (Article 71 par. 3).

The main oversight and enforcement authorities under the EU AI Act are 'market surveillance authorities' (MSAs), which constitute an oversight model that is common in EU product law.[71] The MSAs in question could be new entities or extensions of existing institutions at the national level; they will be responsible for post-market monitoring and investigating AI-related incidents. Regarding XAI, these entities will also have the power to oversee and enforce the AI explainability standards.[72] The need for a supranational oversight body at the EU level will inevitably emerge, as has been the case with anti-money laundering supervision,[73] due to the absence of obstacles to the cross-border movement of AI systems.[74] The European Parliament has correctly proposed to establish such a new EU body, an AI Office responsible for providing guidance and coordinating joint cross-border investigations. This oversight body, operating either at a national or supranational level, must have not only effective powers (audits, inspections and reporting requirements) but also the necessary resources (funding, staff and technology) to ensure ongoing compliance. The need for international cooperation in standardising XAI approaches will also certainly emerge as other jurisdictions develop their regulatory frameworks in addition to the EU. This will create the need for harmonising regulations internationally, promoting global best practices, and

---

[71] Michael Veale and Frederik Zuiderveen Borgesius, 'Demystifying the Draft EU Artificial Intelligence Act' (2021) 4 Computer Law Review International 97.

[72] Nevertheless, the European Data Protection Board (EDPB) and European Data Protection Supervisor (EDPS) highlight that certain aspects of the AI regulation proposal are unclear, such as the roles, powers, and (most importantly) independence of market surveillance authorities; see European Data Protection Board and European Data Protection Supervisor, 'Joint Opinion 5/2021 on the proposal for a Regulation of the European Parliament and of the Council laying down harmonised rules on artificial intelligence' (2021) EDPS-EDPB Joint Opinion.

[73] Georgios Pavlidis, 'The birth of the new anti-money laundering authority: harnessing the power of EU-wide supervision' (2023) Journal of Financial Crime (Latest Articles), <https://doi.org/10.1108/JFC-03-2023-0059> accessed 15 January 2024; Georgios Pavlidis, 'Learning from failure: cross-border confiscation in the EU' (2019) 26 Journal of Financial Crime 683.

[74] Moreover, it will be difficult to identify a single competent authority for an AI operator that is active on several national markets; Martina Anzini, 'The Artificial Intelligence Act Proposal and its implications for Member States' (2021) EIPA Briefing 2021/5.

avoiding conflicts in cross-border AI deployments. The EU must be ready to work in this direction.[75]

Finally, the specific XAI framework needs to be dynamic and able to evolve based on advancements in AI technology and the evolution of the general regulatory framework for AI. Mechanisms must be put in place at the national and EU levels for periodic reviews of and updates to the relevant regulations in order to accommodate new challenges and developments in AI and XAI.

## 6. Fostering Trust and Accountability: The Broader Implications of Explainability in AI

The concept of explainability is not just a legal term and a software/technical challenge. It is a principle with far-reaching societal implications, as it can foster trust and empower the public to engage with AI technologies more actively. Furthermore, explainability is a prerequisite for accountability and fairness since a clear XAI decision-making process allows society to hold an entity responsible for its decisions. Enhancing the availability of explanations makes individuals, enterprises and public institutions more informed about the decision-making processes of AI, thus allowing them to make informed choices about AI applications. Hence, XAI is indispensable for ensuring trustworthy and responsible AI.

Of course, policymakers must balance the need for detailed insights into AI models with the potential cognitive overload that threatens users. This requires a clear regulatory vision, standardised approaches to explainability and the deployment of innovative solutions in information architecture, visualisation, interaction design and education. Moreover, XAI requirements must take into account the concerns of the AI industry, such as the cost of XAI, the trade-off between model transparency and performance, and the protection of proprietary algorithms. As the EU's efforts to regulate this technology demonstrate, developing a standardised approach to AI and explainability proves challenging due to the need to accommodate diverse domains and applications in a rapidly evolving technological landscape. The new EU legal framework for AI must consider this diversity and the need to balance innovation with explainability, accountability and public trust.[76]

---

[75] Peter Cihon, 'Standards for AI Governance: International Standards to Enable Global Coordination in AI Research & Development' (2019) Technical Report, Future of Humanity Institute, University of Oxford; see also Oxford Analytica, 'EU's leadership on AI governance faces tough tests' (2021) Expert Briefings, <https://doi.org/10.1108/OXAN-DB261321> accessed 15 January 2024.

---

[76] Stefan Larsson, 'AI in the EU: Ethical Guidelines as a Governance Tool' in Antonina Bakardjieva Engelbrekt and others (eds), *The European Union and the Technology Shift* (Palgrave Macmillan 2021); Anna Marchenko and Mark Entin, 'Artificial Intelligence and Human Rights: What is the EU's approach?' (2022) 3 Digital Law Journal 43.